\documentclass[%
prb,twocolumn,
superscriptaddress,
 amsmath,amssymb,
 aps,
]{revtex4-1}
\usepackage[utf8]{inputenc}
\usepackage[T1]{fontenc}
\usepackage{mathtools}
\usepackage{subfigure}
\usepackage{graphicx}
\usepackage{dcolumn}
\usepackage{bm}
\usepackage{xcolor}
\usepackage{comment} 
\usepackage[normalem]{ulem}

\usepackage{hyperref}

\begin{document}
\author{Aris Dimou}
\affiliation{Interdisciplinary Centre for Advanced Materials Simulation (ICAMS) and Center for Interface-Dominated High Performance Materials (ZGH), Ruhr-University Bochum, Universit\"atsstr 150, 44801 Bochum, Germany}
\author{Pierre Hirel}
\affiliation{Univ. Lille, CNRS, INRAE, Centrale Lille, UMR 8207–UMET - Unit\'e Mat\'eriaux et Transformations, F-59000 Lille, France}
\author{Anna Gr\"unebohm}
\affiliation{Interdisciplinary Centre for Advanced Materials Simulation (ICAMS) and Center for Interface-Dominated High Performance Materials (ZGH), Ruhr-University Bochum, Universit\"atsstr 150, 44801 Bochum, Germany}

  \title{
  Pinning of domain walls by strontium layer in BaTiO$_3$ perovskite:\\ an atomic-scale study }

  \date{\today}

\begin{abstract}

\ We use atomistic simulations to study the interactions between two-dimensional domain walls and Sr inclusions in the prototypical ferroelectric BaTiO$_3$. Based on nudged elastic band calculations we predict that the energy barrier for domain wall movement increases in the vicinity of small planar Sr inclusions which may act as pinning centers. We link this observation to the local increase in polarization by larger oxygen off-centering and validate our predictions by molecular dynamics simulations of field-driven domain walls at finite temperatures.
  
\end{abstract}

\pacs{ferroelectrics; domain walls; molecular dynamics simulations}

\maketitle

\section*{Introduction}

\ In the family of ferroelectric materials, solid solutions based on barium titanate (BaTiO$_3$ or BTO) combine superior functional responses to applied electric fields, making them suitable for applications such as efficient solid-state cooling devices, piezoelectric energy harvesting, energy storage, and data storage devices based on polarization switching.\onlinecite{acosta_batio_2017, said_ferroelectrics_2017, grunebohm_interplay_2022, yang_perovskite_2019, jiang_enabling_2022}

\ Most functional responses depend on the field-induced motion of domain walls \cite{ damjanovic_ferroelectric_1998, xu_stationary_2014, meier_ferroelectric_2022} which has been studied since decades. Already in the 1950s, Merz reported that the DW velocity in an applied electric field ${\bm E}^{\text{ext}}$ is proportional to $\exp (-{\bm E}^{\text{ext}}_a/{\bm E}^{\text{ext}} )$, where ${\bm E}^{\text{ext}}_a$ is the activation field needed to overcome the activation energy $E_{\text{a}}$ for wall motion. \cite{merz_domain_1954, miller_direct_1959} The microscopic understanding and optimization of the domain wall dynamics are, however challenging. Revisiting DW dynamics with modern laboratory equipment and atomistic simulations showcase new insights: \cite{grunebohm_interplay_2022}

\ Density functional theory has been successfully used to analyze the character of static domain walls, \cite{meyer_ab_2002, grunebohm_domain_2012,li_first-principles_2014} determine activation energies $E_{\text{a}}$  \cite{li_domain_2018} and predict an increase of $E_a$ with an increasing polarization under strain. \cite{beckman_ideal_2009} Molecular dynamics simulations (MD) revealed that domain wall motion is governed by nucleation and growth of 2-D clusters on the moving wall, \cite{liu_intrinsic_2016} that new domains may also nucleate in the center of the domain for larger fields~\cite{boddu_molecular_2017}, or that non-equilibrium dynamics of dipoles may initially boost the DW velocity for ultra-fast field changes. \cite{khachaturyan_domain_2022}

\ The most common route to optimize functional ferroelectrics is substitution.\cite{acosta_batio_2017} In particular, addition of smaller ions such as Sr in the solid solution of (Ba,Sr)TiO$_3$ decreases the volume, the macroscopic polarization and the Curie temperature $T_c$ and thus shifts the maximal functional responses close to the transition to ambient temperatures. \cite{menoret_structural_2002, lemanov_phase_1996, tinte_ferroelectric_2004, nishimatsu_molecular_2016, gruenebohm_optimizing_2018} This sensitivity of material properties to Sr concentration suggests a large impact of concentration gradients and inhomogeneities on the functional properties. Indeed it has been reported that the temperature window with large functional responses \cite{liu_enhanced_2014, damodaran_large_2017} may broaden by inhomogeneities and for the limiting case of  superlattices, novel complex domain morphologies and negative capacitance have been found.\cite{lisenkov_unusual_2009, estandia_rotational_2019, walter_strain_2020}

\ While it is known for other inhomogeneities such as point defects, dislocations, or grain-boundaries that their interaction with domain walls may also be detrimental to applications due to domain wall pinning \cite{yang_direct_1999, jesse_switching_2006, leschhorn_influence_2017, li_domain_2018, pramanick_domains_2012, damjanovic_ferroelectric_1998}, and their role in functional fatigue, \cite{lupascu_fatigue_2005, genenko_mechanisms_2015} the impact of concentration gradients and inclusions is less established. For paraelectric ZnO inclusions in (Bi, Na)TiO$_3$ the observed hardening suggests domain pinning. \cite{kv_hardening_2017, bai_enhanced_2018} Phase-field simulations revealed that paraelectric SrTiO$_3$ layers may pin domain walls in BaTiO$_3$, as it is found that it is energetically more favorable for the DW to be situated on the paraelectric inclusion. \cite{stepkova_pinning_2018} 

\ To the best of our knowledge, a systematic atomistic understanding is so far lacking. This motivates us to investigate atomistically the role of planar SrTiO$_3$ inclusions on the mobility of 180$^\circ$ DWs in tetragonal BTO. We find that ultra-thin inclusions may indeed pin domain walls, however, in contrast to the previous understanding the walls are trapped next to the inclusions while walls inside the inclusion are least favorable.

\begin{figure*}[ht!]
 \centerline{
   \subfigure[]{\includegraphics[width=0.55\textwidth,clip,trim=4cm 9.5cm 3.cm 8.5cm]{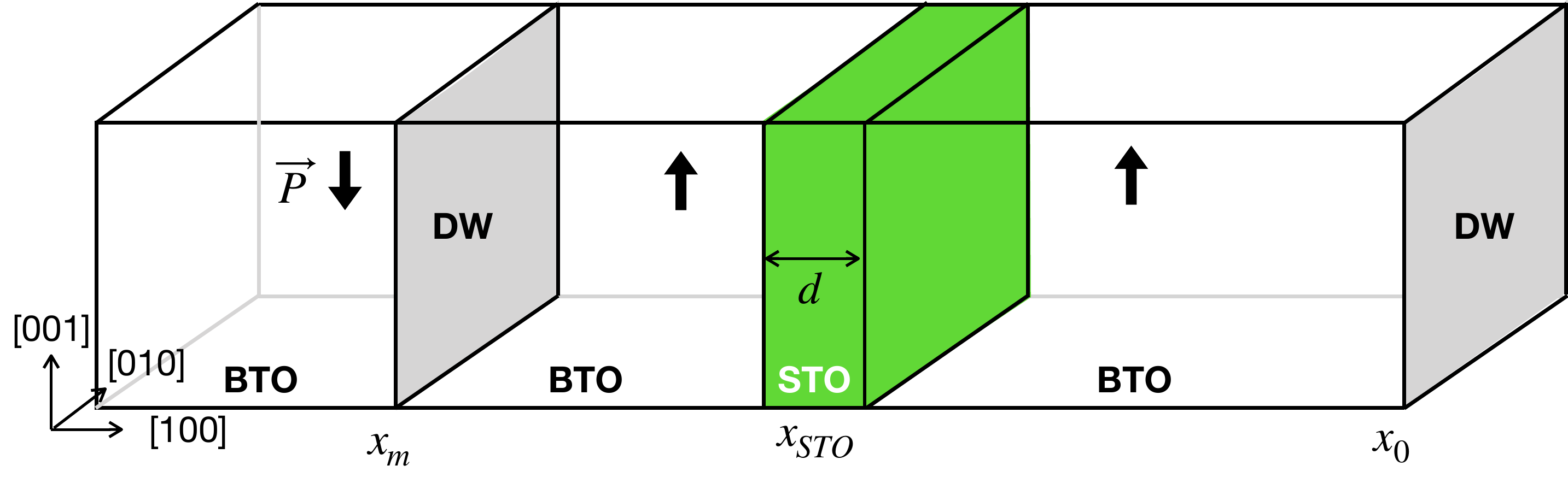}}
    \subfigure[]{\includegraphics[width=0.35\textwidth,clip,trim=1.5cm 5.7cm 8cm 2cm]{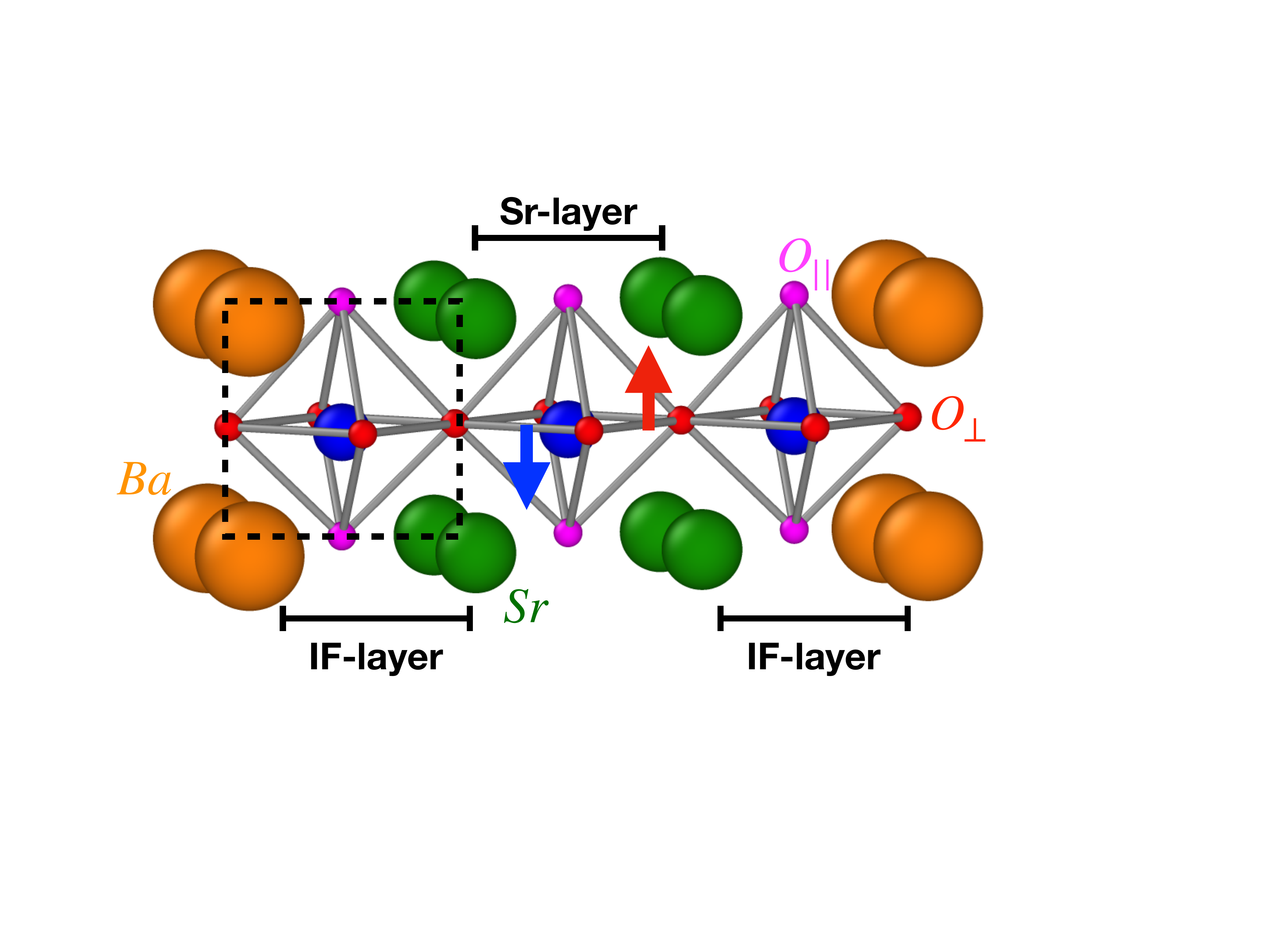}}}
  \caption{
  Schematics of the simulation setup: (a) A $20 \times 3 \times 3$ supercell of BaTiO$_3$ contains domains with electric polarization $\vec{P}$ along $\pm [001]$ (black arrows), separated by 180$^\circ$ domain walls at $x_m$ and $x_0$ (gray planes). A Sr inclusion is centered at $x_\text{STO}$, and walls and inclusions are normal to $[100]$. When an external electric field is applied along $Z=[00\bar{1}]$, DWs move along $X=[\pm100]$. (b) Extract the simulation cell across an inclusion with $d=2$ showing the atomistic structure. An oxygen octahedron surrounds the Ti ion (medium blue spheres). In the ferroelectric phase with polarization along $[00\pm 1]$, these atoms shift with respect to each other (see blue and red arrows). One may distinguish two O$_{||}$ (magenta) and four O$_{\perp}$ (red) atoms with Ti-O bonds either parallel or perpendicular to the polarization direction. The corners of the unit cells are occupied by Ba (orange) or Sr (green) atoms. Interface (IF) layers are defined as separating a SrO layer on one side, and a BaO layer on the other side.
  }
\label{fig:schematic}
\end{figure*}


\section{Methods and models}

\ We model atomic interactions by atomistic pair potential parameterized to reproduce key material properties of the solid solution of (Ba,Sr)TiO$_3$ (BSTO) computed with density functional theory (DFT) by Sepliarsky and co-authors. \cite{sepliarsky_atomic-level_2005} Interactions between ions include Coulomb forces and Buckingham potentials with a cut-off of 12~\AA{} for short-range interactions. Each ion is modeled as a positive core bound to a negative shell, which interacts through an anharmonic spring force \cite{mitchell_shell_1993} to account for the electronic polarizability of ions. 

\ Calculations are performed with LAMMPS, \cite{thompson_lammps_2022} where the Coulomb interactions are computed using the particle-particle particle-mesh method. This potential was shown to reproduce qualitatively the phase diagram of BSTO as well as polarization and structural properties. \cite{tinte_ferroelectric_2004} Furthermore, we verify the observed trends in the local polarization by DFT simulations using the Abinit package. \cite{gonze_abinit_2009, dimou_ab_nodate} The simulation setup is schematically represented in Fig.~\ref{fig:schematic}. The system contains $20 \times 3 \times 3$ unit cells (u.c.) along $X=[100]$, $Y=[010]$, $Z=[001]$, respectively, i.e.\ about $80$~\AA~$\times 11.9$~\AA~$\times 12.3$~\AA{} or 1,800 atoms. Without loss of generality, polarization of the tetragonal BTO phase is initialized parallel to $Z=[001]$, and 180$^{\circ}$ domain walls normal to $X=[100]$ are placed at $x_m$ and $x_0$. The formation energy of a DW can be computed as ${E_{f}} = (E_{p}-E_{DW}) / (2A)$, where $A$ is total wall area in the system, $E_{p}$ and $E_{DW}$ are the total energies of the same configurations with and without a DW.

\ We apply 3-D periodic boundary conditions, thus effectively modeling a periodic array of infinite domain walls in the bulk material. We verified that the chosen dimensions guarantee convergence of the results with respect to system size, i.e.\ the energy barrier for domain wall movement is converged up to 10$^{-4}$ eV/\AA$^2$ and no interactions between neighboring domain walls are expected for this wall distance. \cite{grunebohm_domain_2012,klomp_switching_2022}

\ Each Ti-centered u.c.\ in pristine BTO contains one Ti-atom surrounded by eight Ba atoms, two O$_{||}$ atoms, and four O$_{\perp}$ atoms shared with $w_i$ neighboring cells ($w_{\text{Ba}}=8$, $w_{\text{O}}=2$), as shown in Fig.~\ref{fig:schematic}~(b). 
We monitor the local polarization $\vec{P}_j$ of each  unit cell $j$ by:\cite{sepliarsky_first-principles_2011} 
\begin{equation}
    \label{eq:polarization}
    \vec{P_j}=\frac{1}{V}\sum\limits_{i}\frac{1}{w_i}q_{i}\vec{r_i},
\end{equation} 
where $V$ is the volume of the unit cell and $q_i$ and $r_i$ are the charge  and  the displacement  with respect to the centro-symmetric configuration of atom $i$.
Finally, it is convenient to monitor the average polarization per layer along X, denoted as $\langle P_z \rangle_x$. Each layer has a thickness of one u.c. and contains three planes: one TiO$_2$ plane, and two (Ba,Sr)O planes.

\ We focus on planar inclusions and introduce $d$ planes of SrO parallel to the domain walls at position $x_\text{STO}$, two u.c.\ apart from $x_m$ ($d=1$: one SrO plane, $d=2$: one full SrTiO$_3$ u.c., $d=3$: two full SrTiO$_3$ u.c., $d=4$: three full SrTiO$_3$ u.c.). In the interface layers (IF), each Ti is then surrounded by four Ba and four Sr atoms. We fix the volume to that of pure tetragonal BTO and relax the atomic positions along $Z=[001]$. These constraints correspond to a thin Sr-inclusion in a much larger BTO matrix and are justified by the similar elastic constants of BTO and STO, \cite{piskunov_bulk_2004} making the overall straining of the system mainly depends on the relative fraction of both elements.

\ Using the nudged elastic band (NEB) method\cite{henkelman_climbing_2000} we monitor the energy landscape for the domain wall displacement along $X=[100]$. Nine intermediate images between the initial and final state are used to achieve good accuracy in the NEB calculation.

\ In a second part, we perform molecular dynamics (MD) simulations to study the motion of DWs under an applied electric field. These simulations are performed in the isothermal-isobaric ensemble (NPT), and temperature and pressure are maintained constant by means of a Nos\'e-Hoover thermostat and barostat. To ensure energy conservation, a time step of $0.4$~fs was chosen and we increase the simulation cell to $20 \times 18 \times 18$~u.c.,corresponding to 64,800 atoms to reduce thermal noise. After thermal equilibration for $20$~ps, we instantaneously apply an external electric field along $[00\bar{1}]$, with a magnitude  ${\bm E}^{\text{ext}}$ ranging from $100$ to $600$~kV/cm. We focus essentially on  ${\bm E}^{\text{ext}} = 125$~kV/cm, i.e. within the range where DW motion is determined by propagation. The MD simulation is continued for up to 1000~ps, and the position of the DW is monitored by analyzing snapshots every $6.4$~ps.

\begin{figure}[t]
  \subfigure[]{\includegraphics[width=0.45\textwidth]{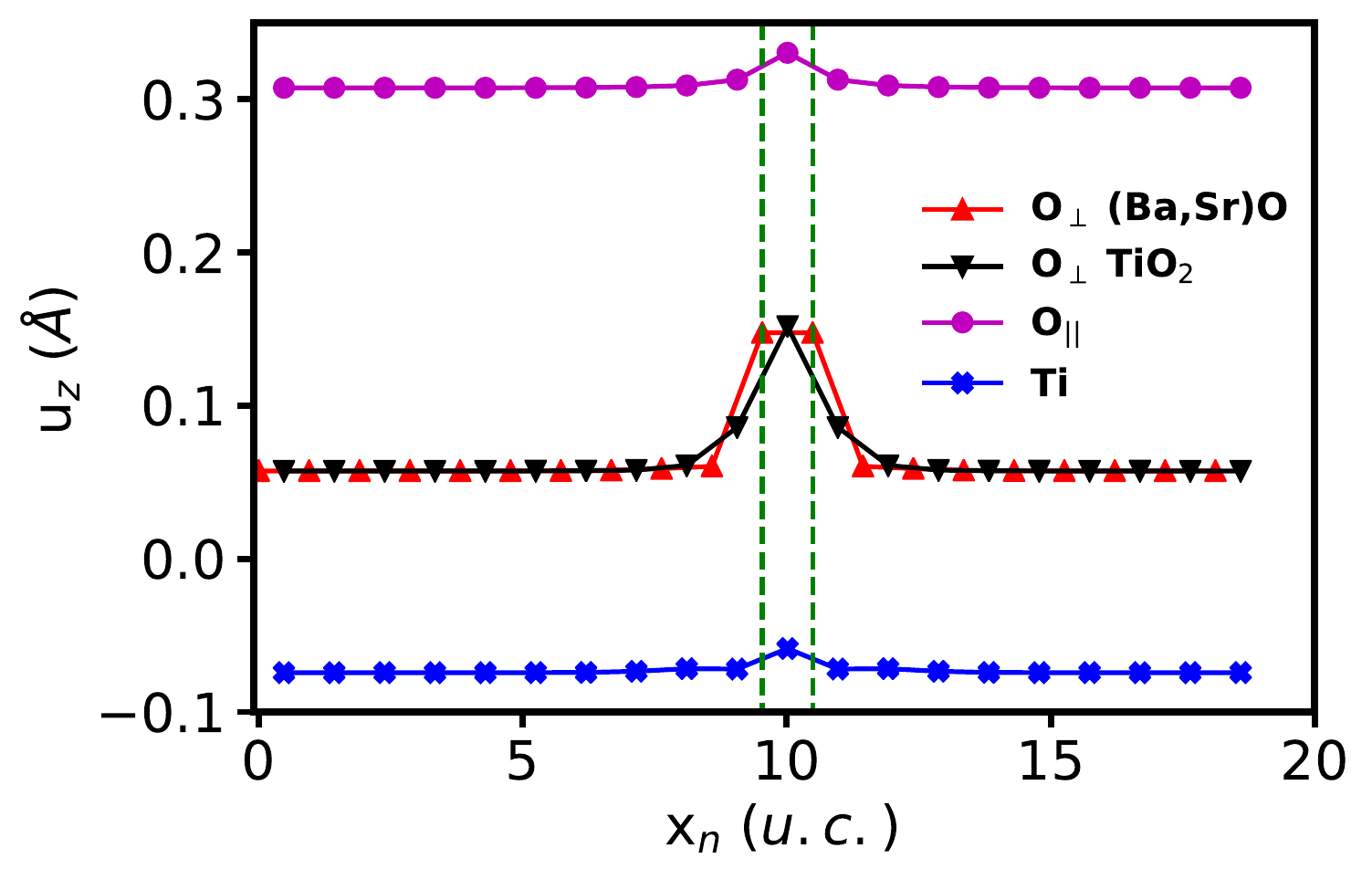}\label{fig:uDis}}
  \subfigure[]{\includegraphics[width=0.45\textwidth]{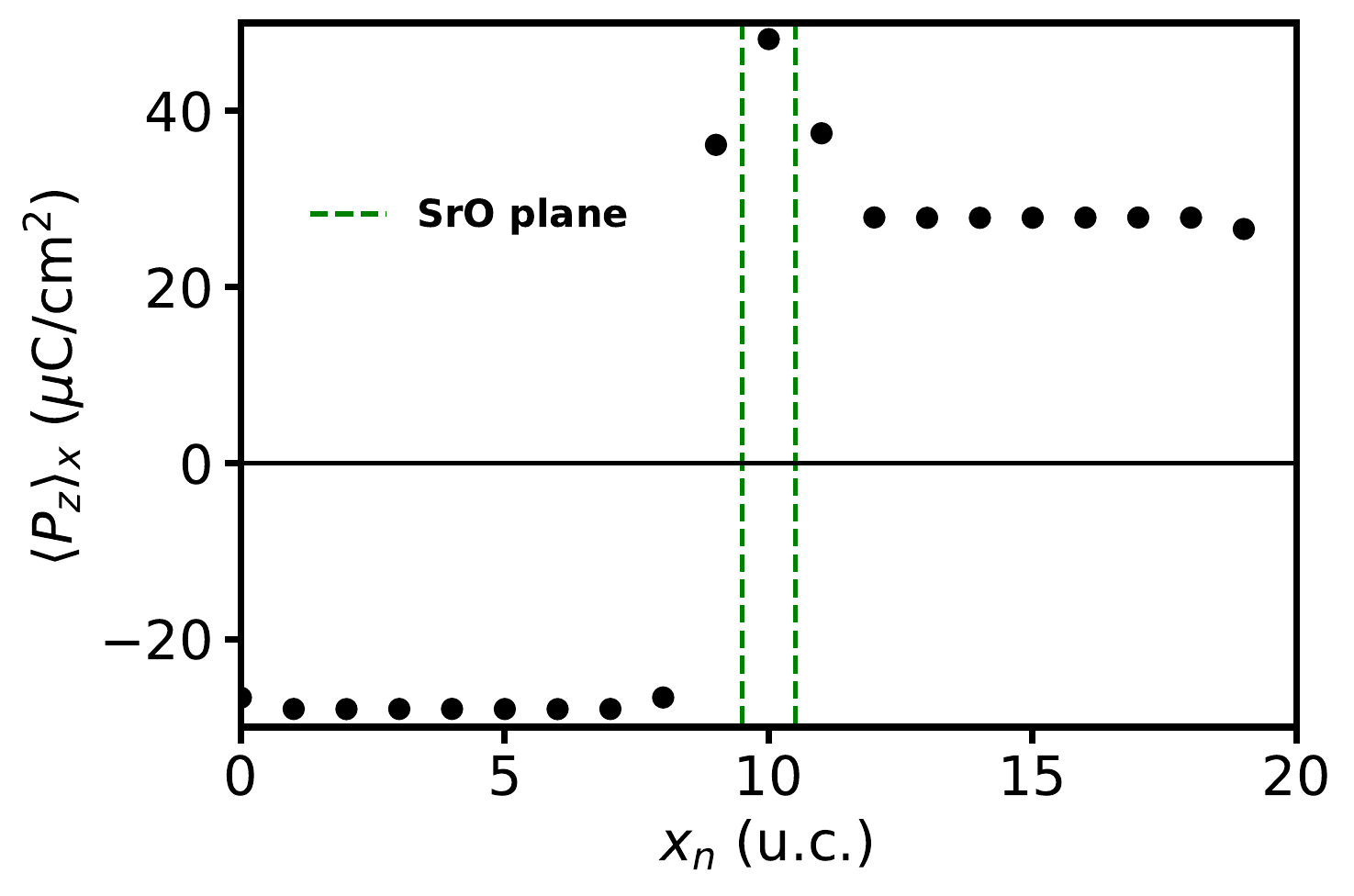}\label{fig:t_pol}}
  \caption{
  Impact of a Sr inclusion with a width of two planes ($d=2$) centered at $x_n=10$ on (a) atomic displacements relative to the centro-symmetric state and (b) polarization profile across a BaO-centered 180$^{\circ}$ DW. Each dot represents the average polarization of unit cells at the given position $x_n$ and vertical black dashed and dashed green lines mark the position of the SrO inclusions. In (a) colors correspond to the atomic displacement of the Ti and O atoms along the Z-axis.
  }
\label{fig:Pol1}
\end{figure}

\section{Atomic structure of the DW}

\ We find that it is most favorable for the 180$^\circ$ DWs in tetragonal BTO to be centered on BaO planes, with a formation energy of ${E_{f}}=0.38$~meV/\AA$^{2}$, in good agreement with \emph{ab initio} calculations \cite{meyer_ab_2002, grunebohm_domain_2012, li_domain_2018}. In addition, we can reproduce the well-known Ising character of the wall, with a width of only 2.1~\AA${}$. \cite{padilla_first-principles_1996, grunebohm_domain_2012} All these results confirm the accuracy of the shell-model potential to describe the ferroelectric behavior of tetragonal BTO.

\ Next, we analyze the impact of a planar Sr inclusion on the local dipole structure and compare it to the surrounding BTO matrix. While the macroscopic polarization in a solid solution is expected to decrease with the Sr concentration, \cite{menoret_structural_2002, nishimatsu_molecular_2016} we find the opposite trend for ultrathin inclusions: it even increases by 30\% at the interface and by nearly 100\% in the pure Sr-layers. This higher local polarization in the vicinity of Sr, however, is only contradicting at first sight and is in agreement with previous atomistic studies. \cite{tinte_ferroelectric_2004,wexler_sr-induced_2019} While the lattice constant in a homogeneous solid solution decreases with the Sr concentration, reducing the overall polarization, the inclusion is strained by the BTO matrix leaving the Ti and O atoms even more space to shift with respect to each other, close to the smaller Sr ions.

\ To get a better understanding of this enhanced local polarization, we compute the displacements of each type of ion with respect to the cubic reference, see Fig.~\ref{fig:uDis} for $d=2$. On the one hand, Sr cations have little impact on the shift of Ti and O ions parallel to the Ti-O bond (O$_{||}$), e.g., on the inclusion, these displacements change by $-2$\% and $+7$\% only. On the other hand, the displacement of oxygen ions perpendicular to the Ti-O bond (O$_{\perp}$) is significantly enhanced in the vicinity of Sr, e.g. (164\%) in both the SrO planes and the TiO$_2$ plane sandwiched between two SrO planes. We note that in contrast to Ref.~\cite{wexler_sr-induced_2019} we do not find an enhanced Ti shift, neither with the used atomistic potentials nor with DFT simulations. 

\ Figure~\ref{fig:t_pol} illustrates the change of the polarization profile if the DW touches a Sr-inclusion. Notably, neither the Ising type of the wall nor the wall width is modified and the polarization of the IF-layer remains higher than in pure BTO even though the DW is situated next to it. We note that although volume relaxation becomes increasingly important with an increasing ratio of Sr to Ba atoms, for $d=4$ inclusion, with a Sr to Ba ratio of 1 in 4, the polarization of the inclusion is still about 28\% larger than in the pristine area. In summary, ultra-thin inclusions of Sr enhance the local polarization. 

\begin{figure*}[t]
  \centerline{
  \subfigure[]{\includegraphics[height=0.28\textwidth]{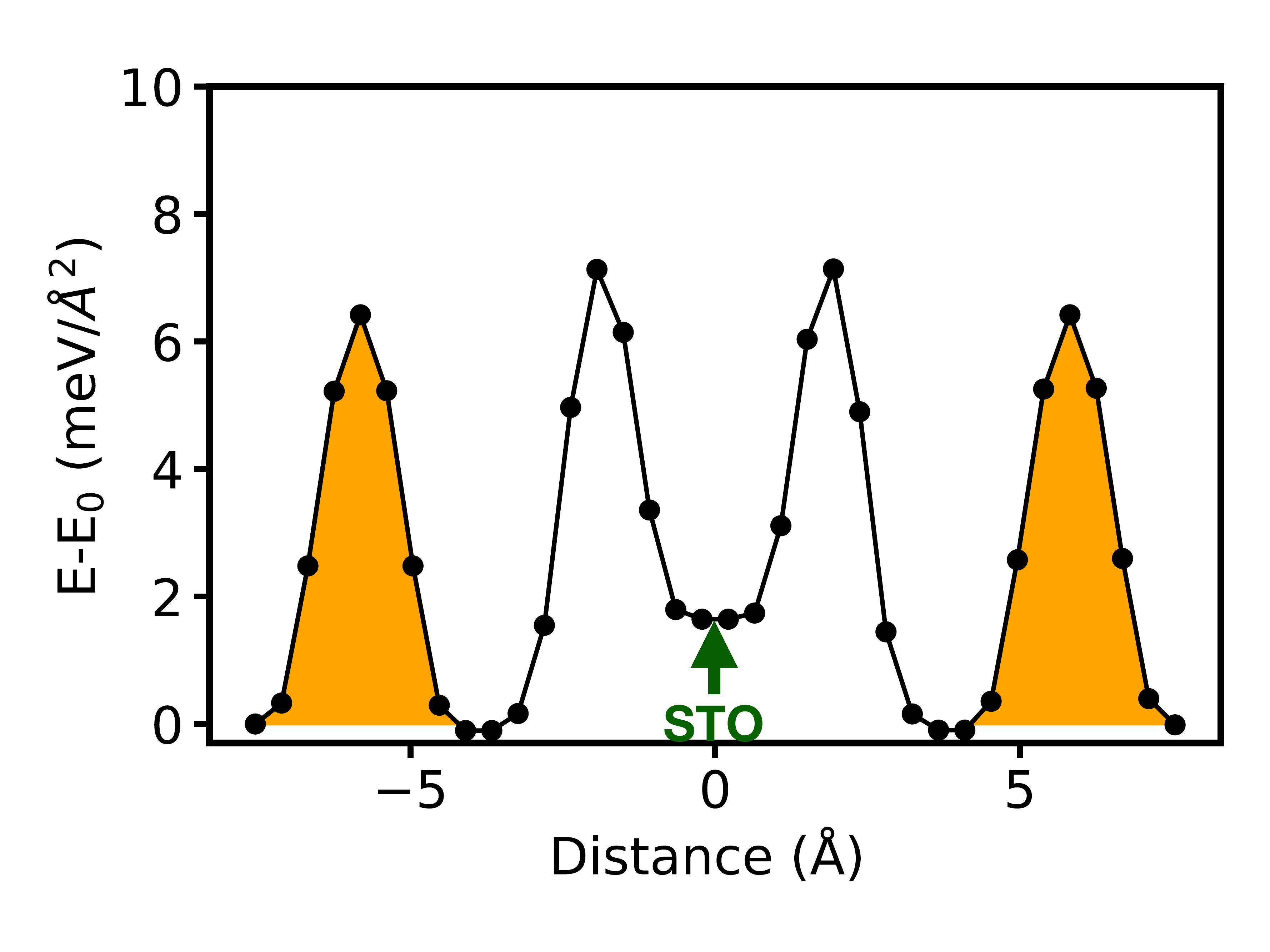}\label{fig:Sr1}}
  \subfigure[]{\includegraphics[height=0.28\textwidth,clip,trim=5.5cm 0cm 0cm 0cm]{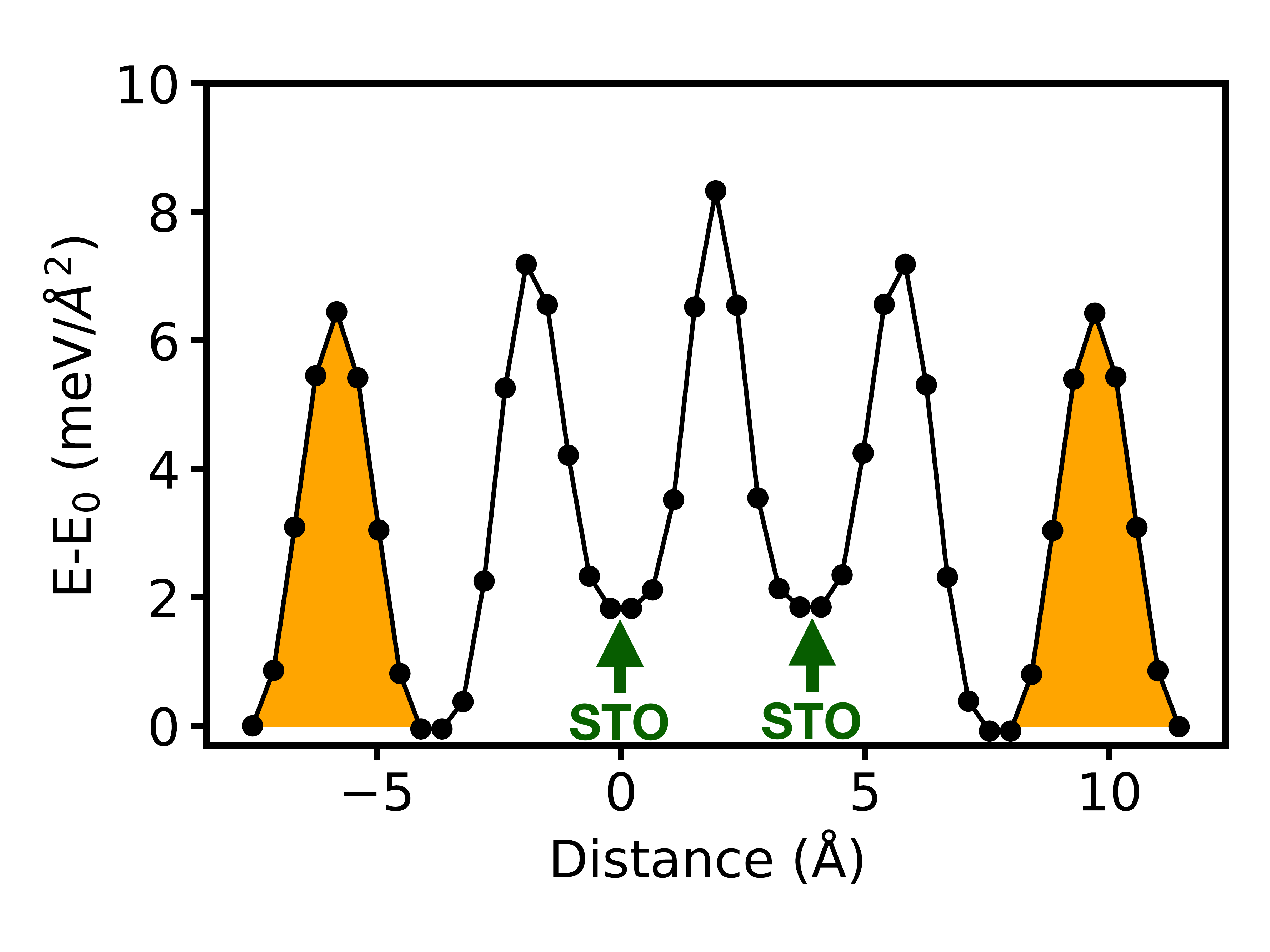}\label{fig:Sr2}}
   \subfigure[]{\includegraphics[height=0.28\textwidth,clip,trim=5.5cm 0cm 0cm 0cm]{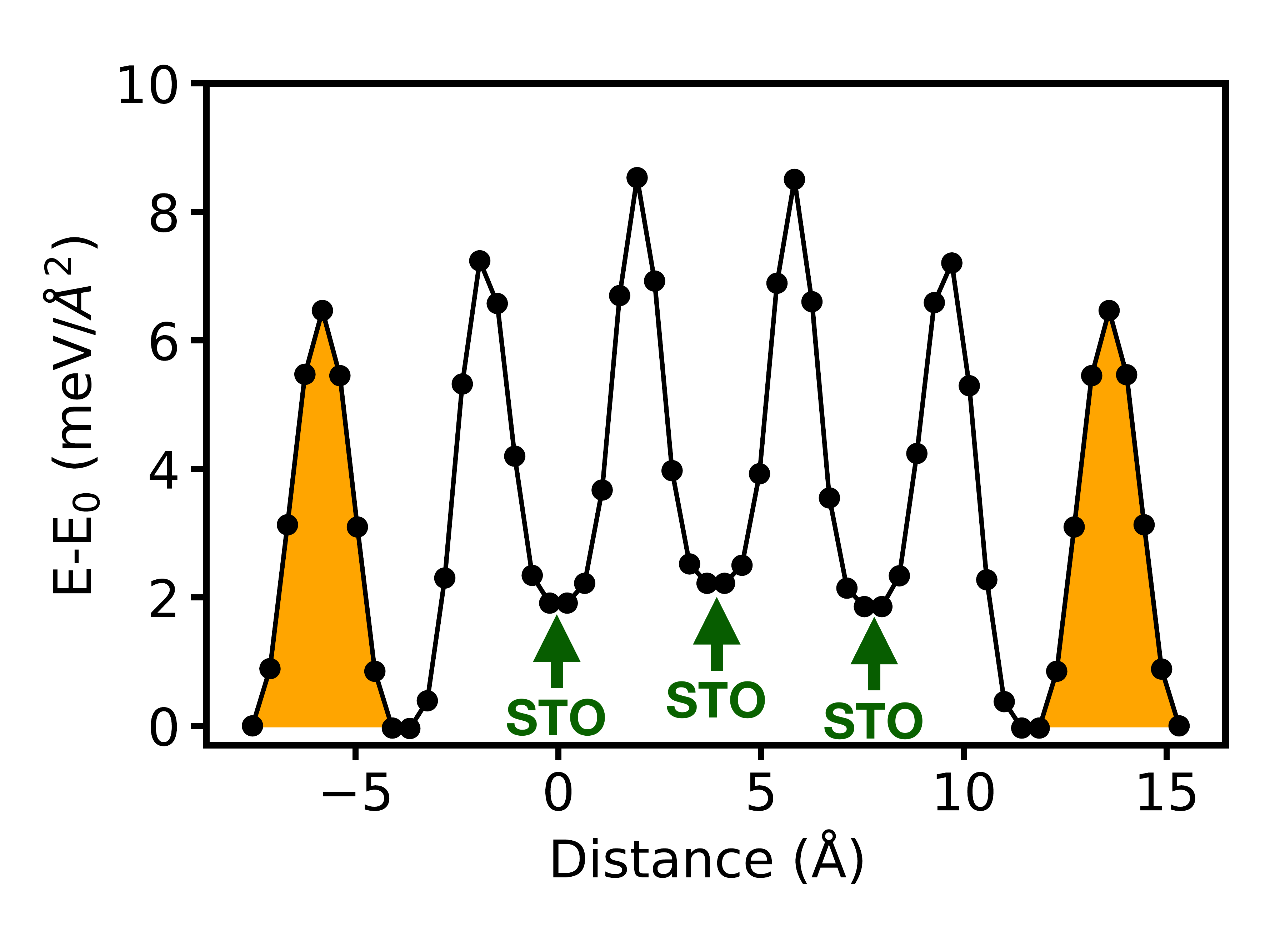}\label{fig:Sr3}}}
  \caption{
  Energy landscape for the static displacement of a 180$^\circ$ DW across a STO inclusion, for (a) a single SrO-plane ($d=1$); (b) two SrO planes ($d=2$); and (c) three SrO planes ($d=3$). The distance is given relative to the first SrO plane on the left of the inclusion, and orange filling indicates pure BTO layers.
  }
  \label{fig:Neb_c}
\end{figure*}

\section{Energy landscape for DW motion}

\ The energy barrier $E_a$ for displacing the domain wall from one local energy minimum to another, i.e., from one BaO plane to the next, is a good indicator of the domain wall mobility. To determine this value we perform NEB calculations with BaO-centered walls as the reference state.

\ The resulting energy landscapes for various widths of inclusions are reported in Fig.~\ref{fig:Neb_c}. In qualitative agreement with previous \emph{ab initio} calculations on pristine BTO \cite{li_domain_2018,grunebohm_domain_2012} we find that TiO$_2$-centered walls are local maxima of energy separating local energy minima on BaO or SrO centered walls, both in the BTO matrix and in the vicinity of Sr. In the BTO matrix the activation energy for DW motion is about $E{_a} = 6.5$~meV/\AA$^{2}$. We note that this value obtained with the shell model is overestimated with respect to previous \emph{ab initio} calculations \cite{meyer_ab_2002, grunebohm_domain_2012, li_domain_2018}, however considering that other key properties are quantitatively reproduced by the shell model, we expect it to give correct qualitative trends.

\ Furthermore, Fig.~\ref{fig:Neb_c} shows that the impact of  Sr is very short-ranged as the energy barrier becomes equal to the one in the pristine material already one unit cell apart from the inclusion. Only in the direct vicinity of Sr atoms does the domain wall formation energy increase. For a DW centred on the central SrO layer, we find a formation energy $E_f$ of $1.43$~meV/\AA$^{2}$, i.e.\ about 2.1~meV/\AA$^{2}$ higher than in pristine BTO.

\ Even more important, the energy barrier for shifting the wall across the TiO$_2$ centered plane in the IF layer increases by 0.7~meV/\AA$^{2}$ compared to that of pure BTO. When related to thermal energy, this corresponds to an increase in temperature of 28~K. Furthermore, the activation energy $E_a$ for the DW motion from one SrO plane to another is 15\% higher than for the motion from a SrO to a BaO plane.

\ These local modifications of the domain wall energy and the energy barrier correlate with the local increase of polarization close to Sr. First, having a larger local polarization next to the domain wall results in a larger energy gradient, thus increasing the domain wall energy. Second, shifting the center of the domain wall to a TiO$_2$ centered plane corresponds to the switching of the polarization on this plane to zero, which is higher in energy in the presence of Sr where the smaller short-range repulsion by smaller ions promotes the ferroelectric instability.

\ Based on the energy landscape for domain wall shifting, several conclusions may be drawn: In pure BTO, the energy landscape is symmetric along $x$, all BaO-centered DW positions being strictly equivalent. In contrast, when Sr atoms are present, the energy landscape becomes asymmetric, and one may expect a higher probability of the DW to leave the inclusion than entering it. When the DW moves under the influence of an external electric field, one has to expect the STO layer to slow down or even pin the DW, depending on the magnitude of the applied field.  


\section{Field-induced domain wall motion at finite temperatures}

\begin{figure*}[t]
  \centering
   \centerline{
   \subfigure[]{\includegraphics[width=0.45\textwidth,clip,trim=0cm 0cm 0cm 0cm]{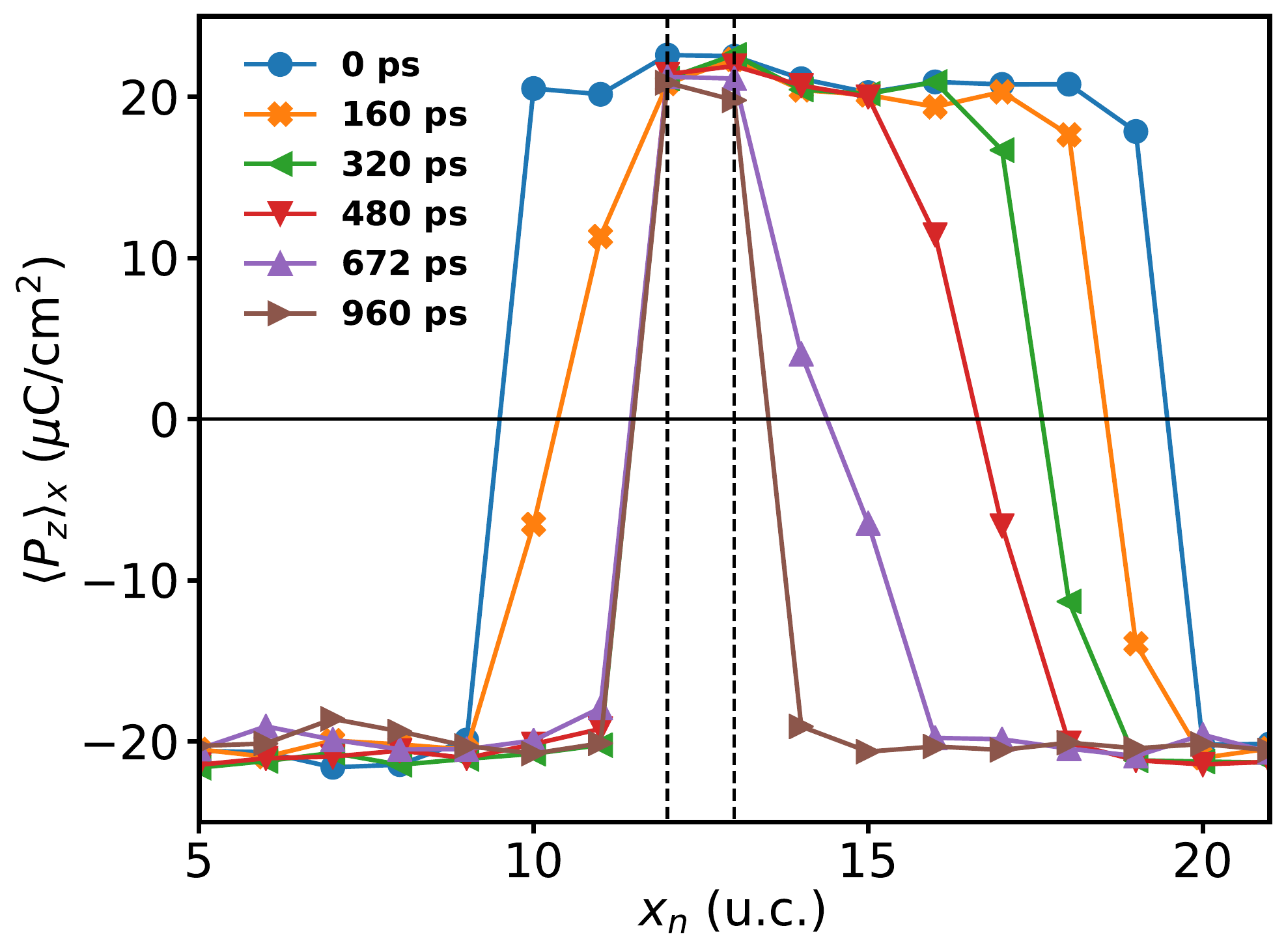}\label{fig:10t}}
  \subfigure[]{\includegraphics[width=0.4\textwidth, clip,trim=5cm 3.4cm 8cm 3cm]{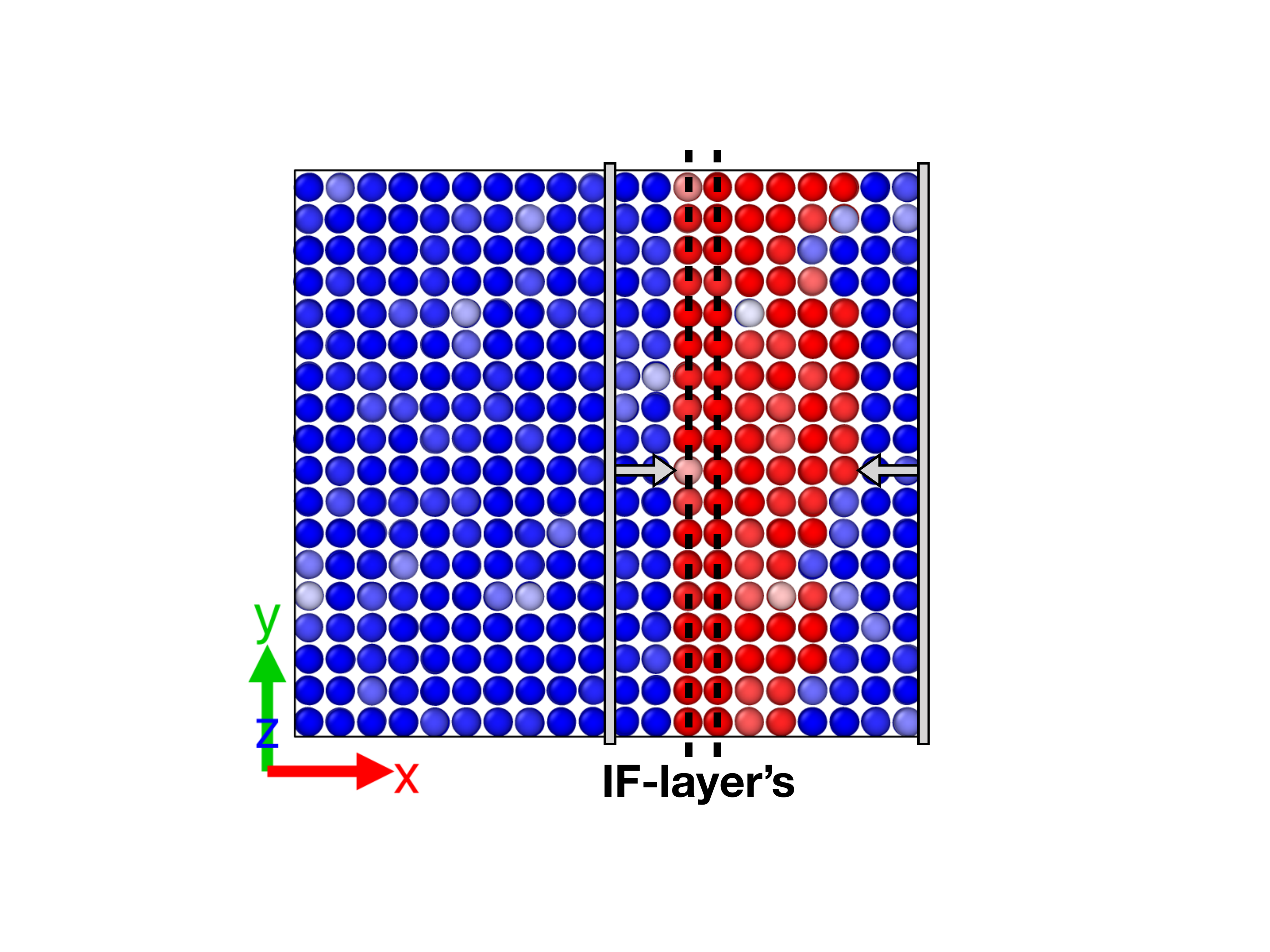}\label{fig:20t}}}
  \caption{
  Field-induced domain wall movement in BTO with a Sr inclusion ($d=1$) between the vertical black dashed lines. Results of Molecular dynamics simulations at 270~K for an electric field of 125~kV/cm applied along -Z. 
  (a)  Change of the layer-resolved polarization with time. 
  (b) Snapshot of the system after 448~ps. Each sphere representing a Ti-centered u.c. is color encoded with the magnitude of the polarization using the blue-white-red color coding with blue being negative, white neutral, and red positive polarization.
  The grey bars show the initial DW positions at $t=0$~ps.
  }
  \label{fig:DW_m}
 
\end{figure*}
\begin{figure}[t]
  \centering
  \subfigure[]{\includegraphics[width=0.47\textwidth,clip, trim=2.5cm 2cm 1.9cm 2cm]{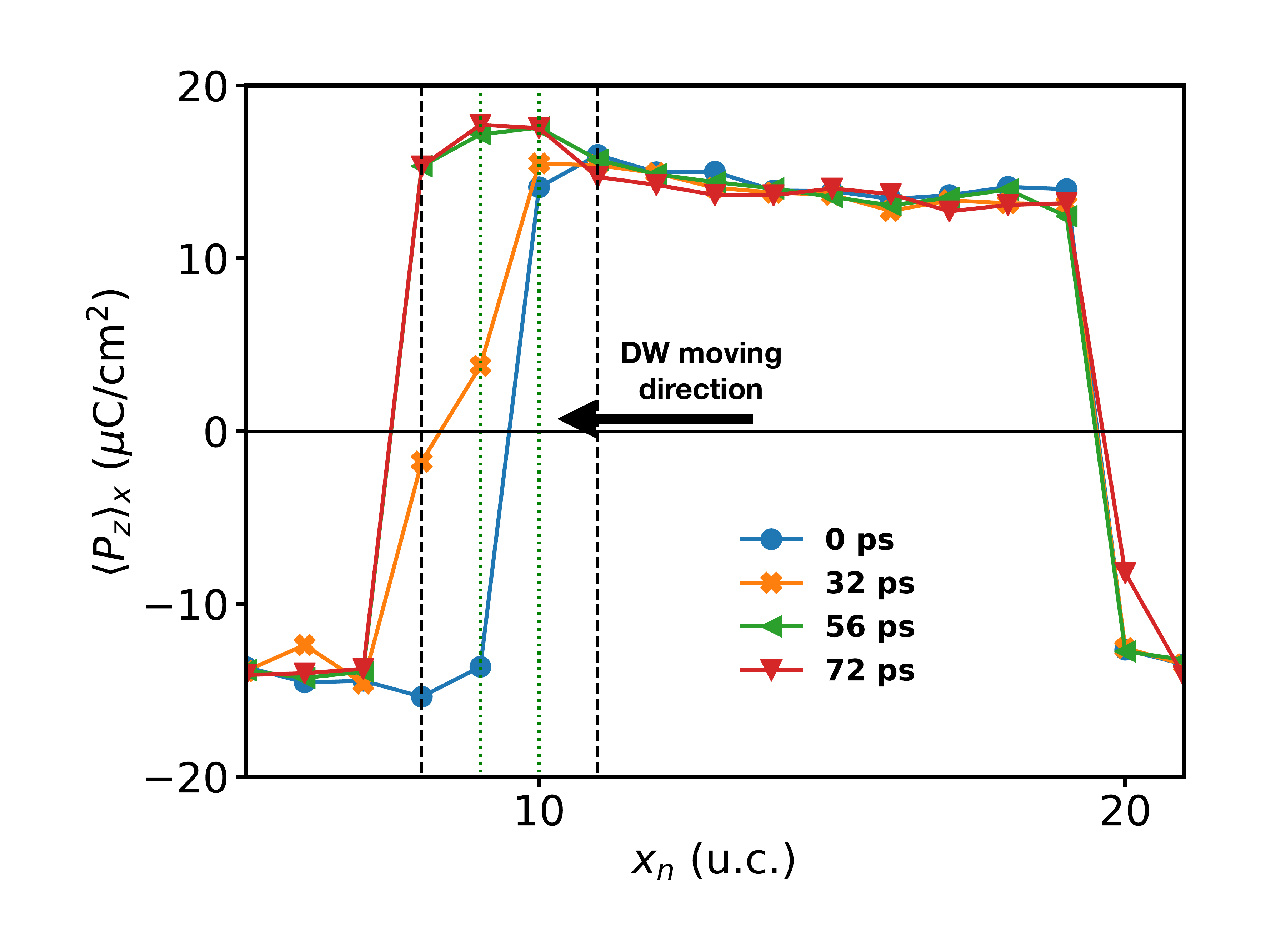}\label{fig:10t}}
  \caption{
  The DW, situated at the tip of the arrow, moves outside the Sr-inclusion at a very fast pace, without the presence of an external applied electric field. The different markers indicate different timesteps in~ps.
  }
\label{fig:Sr_inclDW}
\end{figure}


\ In order to test our hypothesis on the interaction of domain walls and inclusions we study the field-induced domain wall motion in the presence of a SrO plane ($d=1$) performing molecular dynamics (MD) simulations at finite temperature. Initially, the system contains two domain walls (at $x_n=9$ and $x_n=20$). An electric field of magnitude ${\bm E}^{\text{ext}}=125$~kV/cm is applied along $[00\bar{1}$], i.e.\ pointing in the same direction as the polarization of the domain in the left-hand side ($x_n < 9$). Our tests indicate that this field strength is sufficient to trigger the motion of the domain wall in pristine BTO. 

\ Figure \ref{fig:DW_m}~(a) shows the time evolution of the average polarization in each layer across the simulation cell. Initially (blue curve at $t=0$~ps), the polarization profile is qualitatively the same as in molecular static simulations, with the DW width slightly increasing due to thermal fluctuations. With time, both walls move towards the inclusion, effectively increasing the size of the domain polarized along $[00\bar{1}]$.

\ Thereby the walls broaden. This is particularly visible for the DW in the right-hand side between $t=320$~ps (green curve around $x_n = 19$) and $t=672$~ps (purple curve around $x_n=15$). It thus seems that the wall broadens while approaching the inclusion which would contradict the conserved domain wall profiles in the static calculations above. Instead, the broadening is mainly due to the acceleration of the DW under the applied electric field. This can be confirmed by looking at the atomic configuration. Fig.~\ref{fig:DW_m}~(b) shows a snapshot at $t=448$~ps, the polarization being represented by a color code, and DWs are located at red/blue interfaces. At this time frame, the DW on the right-hand side is moving in the field but is still apart from the inclusion. We see that this DW is not planar anymore, but is curved. This is why, when averaging the polarization in each layer along $x$, the DW seems to be broader. This finding is fully in line with previous reports on field-induced domain wall motion in pristine BTO. \cite{shin_nucleation_2007, khachaturyan_domain_2022}

\ Directly at the inclusion, the domain walls become flat again and the mean domain wall width again resembles the static values, see Fig.~\ref{fig:DW_m}.
Note that due to the different initial distances, the left DW reaches the inclusion at $t=320$~ps, while the right DW needs $t=800$~ps. Most important, both DWs are pinned in front of the Sr layer.

\ In other words, the applied field is too small to move the DWs across the  Sr interface, due to the increased energy barrier as predicted by our static calculations. This means that the Sr layer acts as a pinning site. Switching its polarization would require higher temperatures or a stronger electric field. Maintaining temperature constant, we performed MD simulations for different electric fields ranging from $100$ up to $600$~kV/cm. We find that the electric field must be at least 600~kV/cm in order to switch the polarization in the Sr layer. It should be noted that for this field strength, the domain wall movement is no longer the main dominant switching behavior of pristine BTO. Indeed our simulations show homogeneous switching within the whole domain polarized [001], in agreement with previous MD simulations. \cite{boddu_molecular_2017}

\ Finally, we test if the system shows the temperature-driven shift of the domain wall away from the inclusion without any external field we predicted based on the higher energy of domain wall formation on SrO planes in  Fig.~\ref{fig:Neb_c}. For this purpose, we use a system with $d=3$  (15\%~Sr) with the DW initially centered on the central SrO plane, as shown in  Fig.~\ref{fig:Sr_inclDW}. Consistent with the static calculations we observe that the DW is pushed out of the Sr-inclusion after $44.8$~ps and that both walls once in the BTO matrix do not move.

\ In summary, our MD simulations confirm that a thin Sr-inclusion in BTO locally enhances the polarization and thereby deforms the energy landscape for polarization switching which results in a) the pinning of domain walls in front of the inclusion, and thus a higher critical electric field for domain wall movement is needed, b) the instability of domain wall centered on the inclusion.

\section*{Summary and Conclusion}

\ We used atomic-scale simulations to investigate the interactions between 180$^\circ$ DWs and SrTiO$_3$ layers of various widths in BaTiO$_3$. For ultrathin inclusions, we find that the local polarization increases close to Sr inclusion. Therefore it is less favorable for a DW to be centered inside the Sr inclusion than in the BTO matrix and we find that the Sr layer increases the energy barrier for DW motion by at least 15\%, meaning that such inclusions may dampen the wall dynamics or even pin DWs. We were able to confirm our predictions by performing molecular dynamics simulations. At a given temperature, and for an applied field sufficient to move the DWs in BTO, we observed that DWs could not cross the Sr layer and were pinned by it. 

\ In conclusion, our findings suggest that planar Sr can further stabilize domain walls and create dead zones where the presence of domain walls is undesirable. 

\ One may expect similar changes in the energy landscape for the domain wall movement for Sr inclusions of different shapes as well as for concentration inhomogeneities in (Ba,Sr)TiO$_3$ solid solutions. Future studies have to reveal the cross-over from the ultra-thin regime to larger paraelectric inclusions. Note that in inhomogeneities confined to one dimension or non-continuous two dimensions, the bowing effect would be present, while continuous two-dimensional defects as studied here, would be avoided. The pinning for the 2-d inclusion plane could be both achieved, in a very thin Sr inclusion with increased polarization or a paraelectric layer, as was found by Stepkova et al. 

\ Such a feature could be exploited in ferroelectric devices. Given that such structures can be synthesized with techniques employed in superlattice structures, it would be very interesting if experimentalists could actually grow such structures, and confirm -or infirm- our predictions.


\section*{Acknowledgements}

The authors acknowledge financial support guaranteed by the Deutsche Forschungsgemeinschaft (DFG) via the Emmy Noether group GR4792/2, the french embassy for technology and science via the PROCOPE mobility grant DEU-20-42/Code JPE 185DEU0616, and thank Dr. Ruben Khachaturyan for fruitful discussion. 

\bibliographystyle{unsrt}
\bibliography{ref}

\end{document}